# Resonance-induced frequency splitting and evanescent modes at temporal interfaces in elastic metamaterials

Cong Chen[1], Kaijun Yi[1,*], Gengkai Hu[1,2,*]

[1]School of Aerospace Engineering, Beijing Institute of Technology, Beijing 100081, China

[2]Marine Science and Technology Domain, Beijing Institute of Technology, Zhuhai 519088, China

* Corresponding author: kaijun.yi@bit.edu.cn (K.Y.), hugeng@bit.edu.cn (G.H.)

**Abstract:**

Temporal interfaces, defined by abrupt changes in material properties, break temporal translational symmetry and enable wave phenomena fundamentally different from those at spatial interfaces. Unlike spatial scattering, temporal scattering preserves momentum rather than energy, leading to instantaneous frequency shifts governed by the dispersion relations on either side of the interface. Existing studies in elastic media have mainly considered non-resonant materials, and allow only one-to-one frequency conversion across temporal interfaces. Here, we propose temporal interfaces formed by the sudden activation of local resonators in elastic metamaterials, which induces a transition from non-resonant to resonant dispersion. We demonstrate that such interfaces can induce frequency splitting among scattered waves and elucidate how the scattered-wave amplitudes are governed by the weighted modal correlation coefficients and impedances. Moreover, a novel temporal evanescent mode, characterized by spatial stationarity and temporal decay is demonstrated after the interface, which is well explained by the negative effective modulus evaluated at imaginary frequencies. These findings establish a foundational understanding of wave dynamics at temporal interfaces involving resonant materials, open new opportunities for wave manipulation in time-varying solids.

**Keywords**: temporal interface, resonant elastic metamaterial, frequency splitting, temporal evanescent mode

## 1. Introduction

The control of elastic wave propagation in solids is a central topic in solid mechanics, with important implications for vibration and noise mitigation, structural health monitoring, and elastic imaging (Su et al., 2006; Sánchez-Dehesa et al., 2011; Doherty et al., 2013). Over the past two decades, elastic metamaterials have greatly expanded the design space for wave manipulation, enabling phenomena such as elastic bandgaps, subwavelength imaging, mode conversion, cloaking, and topologically protected wave transport ((Liu et al., 2000; Milton et al., 2006; Ambati et al., 2007; Zhu et al., 2014; Zhang et al., 2020; Cheng et al., 2024). These advances predominantly rely on engineered spatial inhomogeneities to steer or confine wave propagation.

Beyond spatial modulation, the mathematical symmetry between space and time in wave equations suggests that elastic waves can also be controlled through temporal or spatiotemporal modulation of material properties. This idea has motivated increasing interest in time-varying elastic media, where phenomena inaccessible to time-invariant systems can emerge such as side frequency generation, wave filtering, amplitude modulation, nonreciprocal wave propagation, mode conversion and waveguiding (Nassar et al., 2017; Yi et al., 2018a; Yi et al., 2018b; Huang and Zhou, 2019; Trainiti et al., 2019; Santini and Riva, 2023; Santini et al., 2024; Cidlinský et al., 2025; Goldsberry et al., 2025) due to the breaking of time-reversal symmetry and energy conservation.

Temporal interfaces, arising from abrupt temporal variations in material properties, differ fundamentally from spatial counterparts. Despite mathematical analogies between spatial and temporal domains (Chen et al., 2021; Wang and Chen, 2024; Wapenaar, 2025), their wave scattering characteristics are distinct. Spatial interfaces conserve energy via temporal continuity, while temporal interfaces preserve spatial translational symmetry and conserve momentum instead (Ortega-Gomez et al., 2023), leading to instantaneous frequency shifts. Causality further restricts scattered waves to propagate only forward in time after interface formation (Pierrat et al., 2025). First explored in electromagnetics (Morgenthaler, 1958; Caloz and Deck-Leger, 2020; Galiffi et al., 2021), temporal interfaces have recently been experimentally realized in elastic systems, demonstrating linear frequency conversion through fast modulation techniques (Kim et al., 2024; Wang et al., 2025).

Despite these advances, studies of elastic temporal interfaces have mainly focused on non-resonant media, where the one-to-one frequency–wavenumber correspondence usually leads only to simple up-or down-frequency conversion. In contrast, resonant media possess multi-valued frequency–wavenumber dispersion and enable much richer spectral responses, as recently demonstrated for electromagnetic waves (Solís and Engheta, 2021; Solís et al., 2021; Hayran et al., 2022; Mirmoosa et al., 2022; Koutserimpas and Monticone, 2024). However, the coupling effect between resonant dispersion and elastic temporal interfaces still remains largely unexploited, as existing elastic investigations are predominantly confined to non-resonant cases (Wang et al., 2025; Chen et al., 2026). Moreover, although temporal scattering in resonant media has been investigated in the context of electromagnetic waves (Solís et al., 2021), the mechanism governing the amplitude distribution among multiple scattered components remains unclear. Furthermore, damping renders resonant media non-Hermitian (Phani, 2022), yet its role in reshaping wave scattering and spectral transformations across non-resonant–to–resonant temporal transitions remains insufficiently understood.

In this study, we will investigate elastic wave dynamics across temporal interfaces of resonant dispersive media generated by the sudden activation of local resonators in elastic metamaterials, using one-dimensional (1D) systems as representative models. The objectives are twofold: (i) to reveal and explain the distinctive wave phenomena induced by the transition from non-resonant to resonant dispersion, and (ii) to determine how damping-induced non-Hermitian features reshapes the temporal-interface response. By considering the instantaneous activation of undamped and damped resonators, respectively, we separately examine the role of resonant dispersion and the influence of dissipation and non-Hermitian effects. Crucially, unlike the one-to-one mode coupling typically found in non-resonant elastic temporal interfaces, the resonant temporal interfaces studied here intrinsically couple a single incident wave to multiple scattered waves. This one-to-multiple coupling lies beyond the scope of conventional temporal-interface theories and necessitates a new framework for prediction and analysis.

The remainder of the paper is organized as follows. Section 2 formulates the 1D elastic metamaterial model with abruptly activated local resonators. Section 3 considers the undamped case and develops a theoretical framework capable of capturing one-to-multiple wave coupling, which is used to predict and analyze the frequency-splitting behavior. Section 4 extends the framework to damped resonators, revealing temporal evanescent modes with real wavenumbers and imaginary frequencies. The origin of these modes is interpreted by extending the conventional concept of negative effective constitutive parameters to the complex-frequency domain, and explicit parameter conditions for their emergence are derived. Finally, Section 5 summarizes the main findings. This work establishes a foundational framework for resonant temporal interfaces in elastic media, and provides physical insight into wave dynamics governed by the interplay between temporal interfaces, resonant dispersion, and non-Hermitian properties.

## 2. One-dimensional elastic metamaterial with abruptly activated local resonators

As a representative model, the 1D elastic metamaterial illustrated in (Fig. 1(a)) is used during the theoretical analysis. The metamaterial contains a mass-spring chain with attached local resonators. The mass in the chain is $m$ and the spring is $k_p$, the distance between two adjacent masses is $a$. The resonators are connected to the masses in the chain through massless rigid rods and hinges, each of them is composed of a mass $m_1$, a spring $k_1$ and a dashpot $c$, which can change with time. The mass $m$ in the chain is constraint to move along the $x$ axis, this horizontal movement is transferred to the vertical direction in the resonators through the hinged rods. Consider the $n^{\text{th}}$ unit cell in the metamaterial, we use $u_n$, $u_{n+1}$ and $v_n^1$ to describe the horizontal and vertical displacements of the masses $m$ and $m_1$, respectively, the vertical displacement of the hinge connecting the two rigid rods is denoted as $v_n^0$. Therefore, the governing equations for waves propagating in the metamaterial with time-varying resonators can be expressed as:

$$m\frac{\partial^2 u_n}{\partial t^2}=\frac{k_1(t)}{\gamma}\left(v_{n-1}^0-v_{n-1}^1\right)/2-\frac{k_1(t)}{\gamma}\left(v_n^0-v_n^1\right)/2+k_p\left(-2u_n+u_{n+1}+u_{n-1}\right) \tag{1a}$$

$$\frac{\partial\left(m_1(t)\frac{\partial v_n^1}{\partial t}\right)}{\partial t}=k_1(t)\left(v_n^0-v_n^1\right)-c(t)\frac{\partial v_n^1}{\partial t} \tag{1b}$$

$$u_n-u_{n+1}=2\gamma v_n^0 \tag{1c}$$

in which, $\gamma = tan\theta$, $\theta$ is the angle between a rigid rod and the $x$-axis.

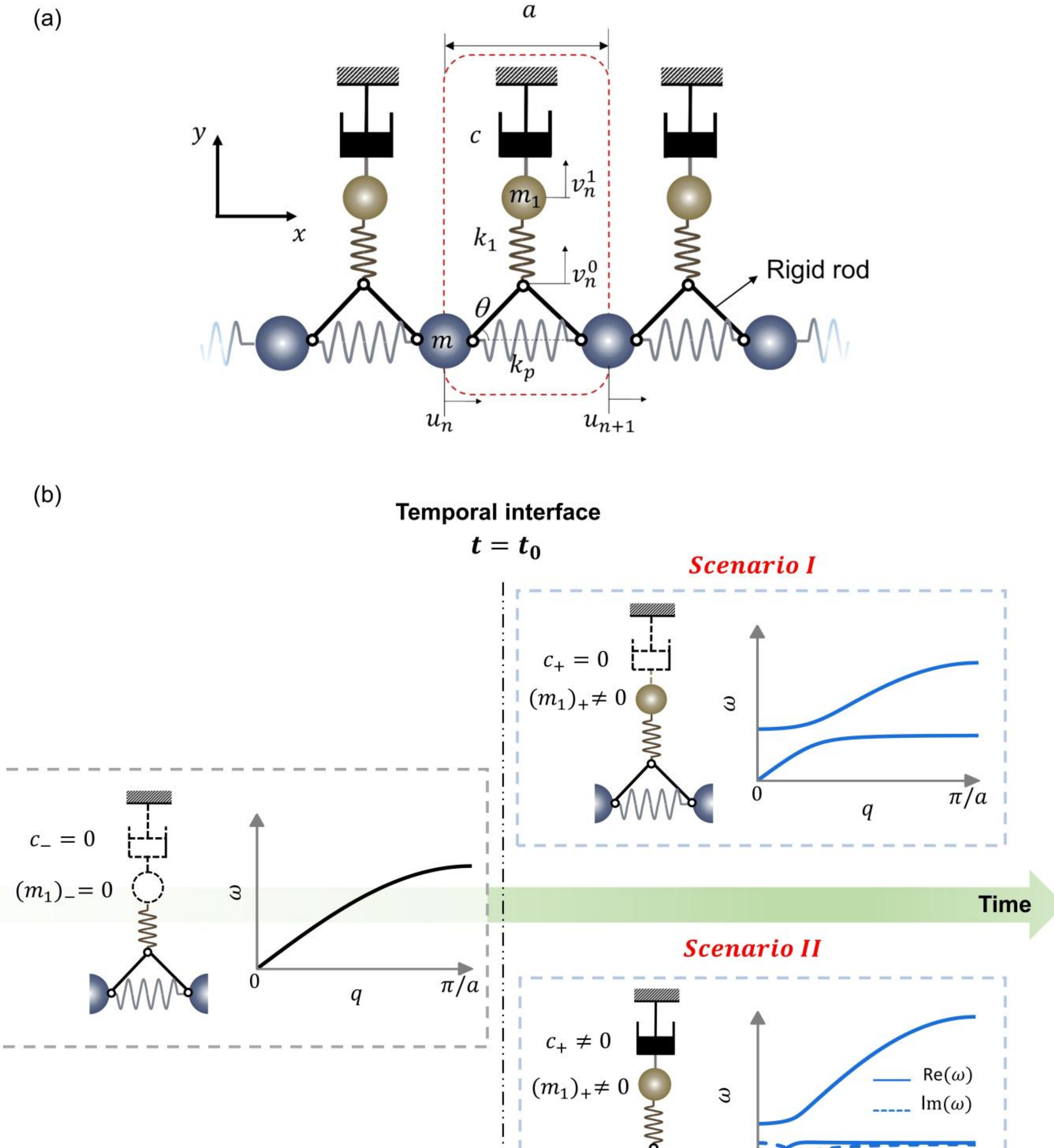

Fig. 1. (a) An infinite one-dimensional locally resonant metamaterial model. (b) A schematic illustration of the abrupt changes of resonator parameters at $t_0$ and the corresponding dispersion curves before and after the interface; two scenarios will be considered, in *Scenario I*, only value of $m_1$ is switched from zero to nonzero, in *Scenario II*, both the values of $m_1$ and *c* are changed to nonzero; $(\cdot)_-$ and $(\cdot)_+$ denote parameters before and after $t_0$, respectively.

To create a temporal interface through the instantaneous activation of local resonators in the metamaterial, several methods are available, for instances, a sudden change either in the resonator mass, from zero to a finite value, or in the resonator stiffness. Different activation protocols lead to different temporal boundary conditions (TBCs), and affect the amplitudes of the temporally scattered waves. However, the qualitative features of the wave phenomena reported in this work are governed by the dispersion relations before and after the temporal interface and therefore are independent of the specific activation protocol. In this study, temporal interfaces generated by activating resonator mass are mainly analyzed for presenting the methodologies, results and analysis. Accordingly, we focus on two representative scenarios illustrated in Fig. 1 (b) ($(\cdot)_-$ and $(\cdot)_+$ represents a parameter just before and after an interface, respectively):

*Scenario I*: before the interface ($t < t_0$), resonators' parameters satisfy $(m_1, c)_- = (0, 0)$, which means the resonators are deactivated. At the interface ($t = t_0$), mass $m_1$ in all resonators is instantaneously changed from zero to a non-zero value $(m_1)_+$.

*Scenario II*: at $t = t_0$, we both change mass $m_1$ and damping $c$ from zero to their respective non-zero values $(m_1)_+$ and $c_+$.

For completeness, Appendix A presents the results obtained by switching the resonator stiffness.

Furthermore, to make our theory and findings more general, we define dimensionless stiffness $\kappa = k_1/k_p$ and mass $\delta = m_1/m$, as well as dimensionless time $\tau = t(\omega_0)_+ \sqrt{\delta_+/\kappa}$ and angular frequency $\Omega = (\frac{\omega}{(\omega_0)_+})\sqrt{\kappa/\delta_+}$. Governing equations in Eqs. (1a)-(1c) can be rewritten using these dimensionless parameters:

$$\frac{\partial^2 u_n}{\partial \tau^2} = \frac{\kappa}{\gamma}\left(v_{n-1}^0 - v_{n-1}^1\right)/2 - \frac{\kappa}{\gamma}\left(v_n^0 - v_n^1\right)/2 + \left(-2u_n + u_{n+1} + u_{n-1}\right) \tag{2a}$$

$$\frac{\partial\left(\delta(\tau)\dfrac{\partial v_n^1}{\partial \tau}\right)}{\partial \tau} = \kappa\left(v_n^0 - v_n^1\right) - 2\zeta(\tau)\sqrt{\kappa\delta(\tau)}\frac{\partial v_n^1}{\partial \tau} \tag{2b}$$

$$u_n - u_{n+1} = 2\gamma v_n^0 \tag{2c}$$

According to the above equations, we can obtain the dispersion relations of the metamaterial before and after the interface:

$$\Omega^2 = 4\sin^2\frac{qa}{2}, \tau < \tau_0 \tag{3a}$$

$$\Omega^2 = 4\left(1+\frac{\kappa}{4\gamma^2}(1-\frac{\Omega_0^2}{-\Omega^2+\Omega_0^2+2i\zeta\Omega\Omega_0})\right)\sin^2\frac{qa}{2}, \tau > \tau_0 \tag{3b}$$

here, $i = \sqrt{-1}$, $\Omega_0 = \sqrt{\kappa/\delta_+}$ is the dimensionless resonance frequency of the resonators after the interface, and $q$ is wavenumber, $a$ is the lattice constant.

In the following, we assume a monochromatic wave with frequency $\Omega_-$, wave number $q$ and amplitude $A_0$ is propagating in the chain before a temporal interface. Without any loss of generality, we set $\kappa = 1$ and $\gamma = 0.5$, values of $\Omega_0$, $\Omega_-$ and $\zeta$ will be specified when necessary.

## 3. Frequency-splitting phenomenon

To elucidate the wave phenomena induced by the transition from non-resonant to resonant dispersion, this section analyzes wave scattering at a temporal interface generated by the activation of undamped resonators, namely, the *Scenario I*. A theoretical framework for the temporal interface is established, encompassing frequency splitting and the distribution of the amplitudes of the scattered waves.

### *3.1 Theory and characteristics of frequency splitting*

#### *3.1.1 Frequencies of the temporally scattered waves*

Since the metamaterial is assumed to be spatially homogeneous before and after a temporal interface, waves across the interface will conserve momentum (Ortega-Gomez et al., 2023), i.e., their wave numbers keep unchanged. With help of this condition, the analytical expressions for the frequencies of the scattered waves can be derived.

Using Eqs. (3a) and (3b), the wave number before and after the interface is respectively derived as $q(\Omega_-) = 2\arcsin\left(\frac{\Omega_-}{2}\right)/a$ and $q(\Omega_+) = 2\arcsin\left(\sqrt{\frac{\Omega_+^2}{4(1+\frac{\kappa}{4\gamma^2}(1-\frac{\Omega_0^2}{-\Omega_+^2+\Omega_0^2}))}}\right)/a$, here $\Omega_+$ is the frequency of a wave after the interface. According to the conservation of momentum, we have $q(\Omega_+) = q(\Omega_-)$, which leads to:

$$\frac{\Omega_+^2}{\Omega_-^2} = 1+\frac{\kappa}{4\gamma^2}\left(1-\frac{\Omega_0^2}{-\Omega_+^2+\Omega_0^2}\right) \tag{4}$$

Solving the above equation yields four roots:

$$\frac{\Omega_+}{\Omega_0} = \pm\sqrt{\frac{\left(1+\left(\frac{\kappa}{4\gamma^2}+1\right)\left(\frac{\Omega_-}{\Omega_0}\right)^2\right) \pm \sqrt{\left(1+\left(\frac{\kappa}{4\gamma^2}+1\right)\left(\frac{\Omega_-}{\Omega_0}\right)^2\right)^2 - 4\left(\frac{\Omega_-}{\Omega_0}\right)^2}}{2}} \tag{5}$$

Obviously, $\Omega_+$ and $\Omega_-$ can be normalized by dividing by $\Omega_0$. Therefore, we set $\Omega_0 = 1$ in this section during the analysis.

The four roots can be categorized into two sets: $\Omega_{F,1}$, $\Omega_{B,1}$ and $\Omega_{F,2}$, $\Omega_{B,2}$ (with $\Omega_{F,1} = -\Omega_{B,1}$, $\Omega_{F,2} = -\Omega_{B,2}$, and $\Omega_{F,1} > \Omega_{F,2}$). Besides, $\Omega$ with subscript F or B have positive or negative real part, it is related to a forward (i.e., time refracted) or a backward (i.e., time reflected) wave in space, respectively. Therefore, Eq. (5) indicates that the incident wave with frequency $\Omega_-$ is transformed to two groups of waves after the temporal interface, and each group is composed of a forward and a backward wave with the same frequency.

To understand the mechanism behind this frequency-splitting phenomenon, we calculate the dispersion curves of the media across the interface, as shown in Fig. 2(a). We assume that a wave with $\Omega_- = 1$ is incident at the interface. Of course, one can calculate the frequencies of the scattered waves using Eq. (5). In addition to this, the momentum conservation condition at the interface can be graphically illustrated by the vertical line in Fig. 2(a). Therefore, the frequencies can also be determined from the intersection points of the vertical line with the dispersion curves for $\tau > \tau_0$ (solid and dashed lines), yielding $\Omega_{F,1} = -\Omega_{B,1} = 1.62$ and $\Omega_{F,2} = -\Omega_{B,2} = 0.62$. From the above analysis, it is clear that the frequency splitting is due to the one-to-two correspondence between frequency and wavenumber in the dispersion relation of the locally resonant metamaterial. Exploiting this mechanism, one can expect more complex wave conversion phenomena by engineering the nonlinearity in dispersion relations.

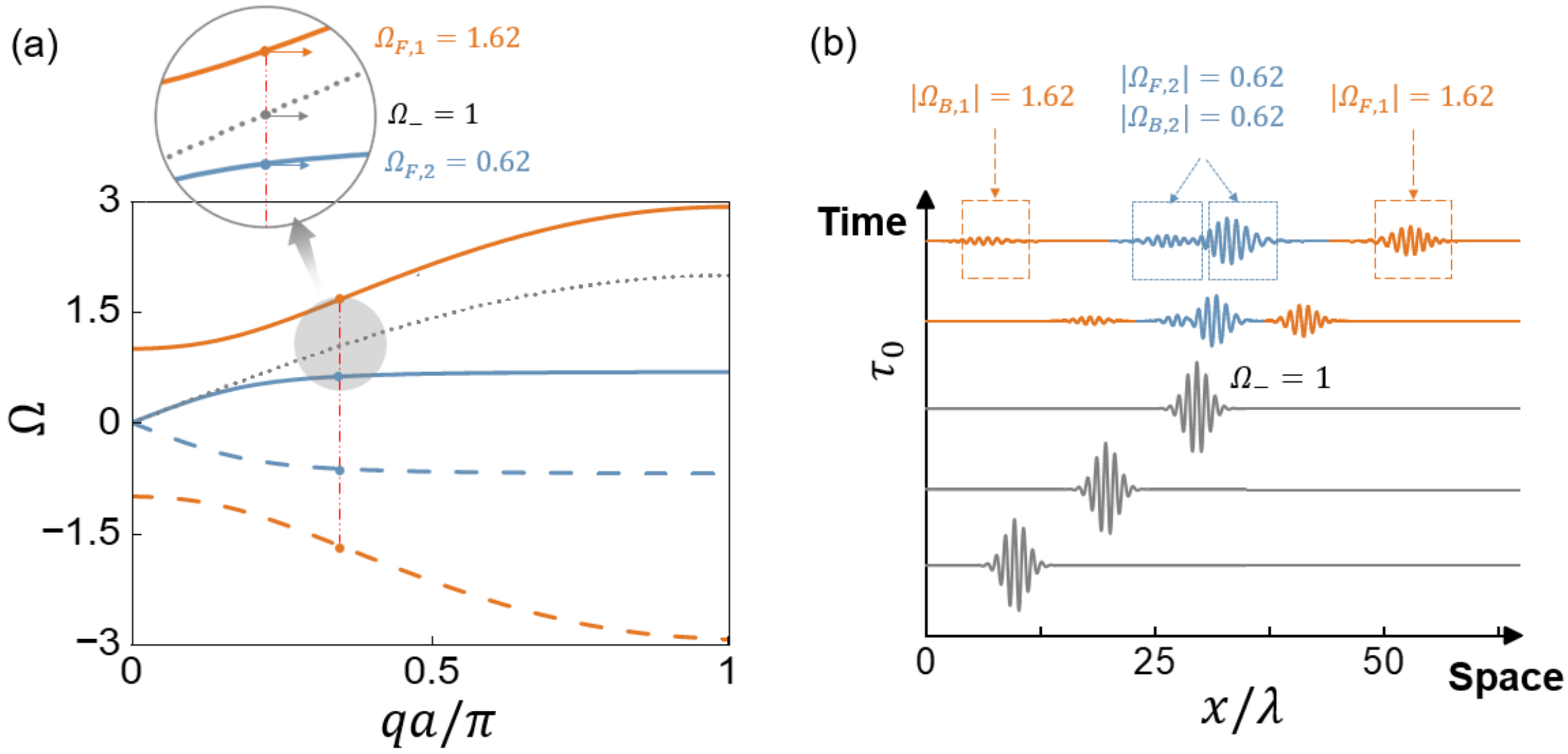

Fig. 2. (a) Dispersion curves before and after the temporal interface ($\tau = \tau_0$), the dotted line represents the dispersion curve for $\tau < \tau_0$, while the other two solid lines and two dashed lines are dispersion curves for $\tau > \tau_0$, the vertical dash-dot line indicates the momentum conservation condition at the interface. (b) Propagation of wave packet before and after the interface ($\tau = \tau_0$) simulated using the FDTD method, $x$ is the spatial coordinate, and $\lambda$ is the wavelength of the incident wave at $\Omega_- = 1$.

To verify the accuracy of the above theory, we conduct numerical simulations using the FDTD method, the results are shown in Fig. 2(b). In the simulation, to be consistent with the theoretical study in Fig. 2(a), we assume that at $\tau = \tau_0$, the mass is instantaneously changed from $\delta_- = 0$ to $\delta_+ = 1$. A 5-cycle tone-burst signal with a central frequency of $\Omega_- = 1$ is generated as incident wave. Multiple scattered wave packets by the interface are clearly observed in Fig. 2(b), central frequencies of them are $|\Omega_{F,1}| = |\Omega_{B,1}| = 1.62$ and $|\Omega_{F,2}| = |\Omega_{B,2}| = 0.62$, in very good agreement with the theoretical predictions in Fig. 2(a).

### 3.1.2 *Amplitudes of the temporally scattered waves*

To uniquely determine the amplitudes of the scattered waves, four independent temporal boundary conditions (TBCs) at the interface must be imposed. Similar to the case without local resonators (Wang et al., 2025), the continuity of displacement and momentum of mass $m$ in the horizontal chain provides two TBCs:

$$u_n \,|_{\tau=\tau_0^-} = u_n \,|_{\tau=\tau_0^+} \tag{6a}$$

$$\dot{u}_n|_{\tau=\tau_0^-} = \dot{u}_n|_{\tau=\tau_0^+} \tag{6b}$$

The above two TBCs are enough to solve the temporal scattering problem for the cases without resonators, but when local resonators are introduced, two extra TBCs are needed.

Mathematically speaking, the first derivatives of $\delta(\tau)\partial v_n^1/\partial\tau$ and $v_n^1$ with respect to $\tau$ must be finite in Eq. (2b). Therefore, the displacement and momentum of mass $m_1$ must also be continuous, giving the two extra TBCs:

$$v_n^1|_{\tau=\tau_0^-} = v_n^1|_{\tau=\tau_0^+} \tag{7a}$$

$$\delta_-\dot{v}_n^1|_{\tau=\tau_0^-} = \delta_+\dot{v}_n^1|_{\tau=\tau_0^+} \tag{7b}$$

Using the TBCs in Eqs. (6a)-(7b), we can derive a temporal scattering matrix to calculate the amplitudes of the reflected and refracted waves.

The displacement field before and after the interface can be expressed as:

$$u_n = A_0 e^{i\Omega_-\tau - iqna}, \tau < \tau_0 \tag{8a}$$

$$u_n = \sum_{m=1}^{2} F_m e^{i\Omega_{F,m}\tau - iqna} + \sum_{m=1}^{2} B_m e^{i\Omega_{B,m}\tau - iqna}, \tau > \tau_0 \tag{8b}$$

We define $\mathbf{A}_+ = [F_1 \quad F_2 \quad B_1 \quad B_2]^{\mathrm{T}}$, $\mathbf{\Omega}_+ = [\Omega_{F,1} \quad \Omega_{F,2} \quad \Omega_{B,1} \quad \Omega_{B,2}]^{\mathrm{T}}$ and $\mathbf{A}_- = [A_0 \quad 0 \quad 0 \quad 0]^{\mathrm{T}}$. Using Eqs.(8a)- (8b) and enforcing the continuity conditions at $\tau = \tau_0$ leads to the following relation:

$$\mathbf{A}_+ = \mathbf{S}\mathbf{A}_- \tag{9}$$

in which, $\mathbf{S}$ is the temporal scattering matrix:

$$\mathbf{S} = \mathbf{P}^{-1}\mathbf{M}_s^{-1}\mathbf{a} \tag{10}$$

with $\mathbf{P} = \mathrm{diag}[e^{i\Omega_{F,1}\tau_0} \quad e^{i\Omega_{F,2}\tau_0} \quad e^{i\Omega_{B,1}\tau_0} \quad e^{i\Omega_{B,2}\tau_0}]$, $\mathbf{a} = \mathrm{diag}[e^{i\Omega_-\tau_0} \quad \Omega_- e^{i\Omega_-\tau_0} \quad e^{i\Omega_-\tau_0} \quad 0]$, and

$$\mathbf{M}_s = \begin{bmatrix} 1 & 1 & 1 & 1 \\ \Omega_{F,1} & \Omega_{F,2} & \Omega_{B,1} & \Omega_{B,2} \\ \dfrac{\Omega_0^2}{-(\Omega_{F,1})^2 + \Omega_0^2} & \cdots & \cdots & \dfrac{\Omega_0^2}{-(\Omega_{B,2})^2 + \Omega_0^2} \\ \dfrac{\Omega_{F,1}\Omega_0^2}{-(\Omega_{F,1})^2 + \Omega_0^2} & \cdots & \cdots & \dfrac{\Omega_{B,2}\Omega_0^2}{-(\Omega_{B,2})^2 + \Omega_0^2} \end{bmatrix} \tag{11}$$

Once $\mathbf{\Omega}_+$ is determined through Eq. (5), the amplitudes of scattered waves can be calculated using Eqs. (9)-(11).

It should be noted that although the analytical framework presented above is derived under the assumption of resonator-mass activation, it is also applicable to temporal interfaces generated by the instantaneous activation of resonators via other mechanisms, only the specific form of the TBCs (Eqs. (7a) and (7b)) needs to be modified.

### *3.1.3 Temporal scattering of incident waves with different frequencies*

Using the above theories, we analyze characteristics of the scattered waves of incident waves with different frequencies. The incident frequency $\Omega_-$ is swept from near-zero to $\Omega_{\text{cutoff}} = 2$, the theoretical results are shown in Fig. 3 together with FDTD results for comparison.

Fig. 3(a) and (b) clearly show that the frequencies and amplitudes of the scattered waves strongly depend on the incident frequency. When $\Omega_-$ is small, far below the resonant frequency $\Omega_0 = 1$, amplitude of the wave $\Omega_{F,2}$ is close to $A_0$ and the other wave components almost vanish, indicating that the temporal interface is nearly transparent. As $\Omega_-$ increases, amplitude of the wave $\Omega_{F,2}$ decreases, amplitudes of other waves increase. Interestingly, when $\Omega_-$ is much larger than $\Omega_0$, the frequency-splitting phenomenon is still very obvious, in contrast to the cases when $\Omega_-$ is much smaller than $\Omega_0$. For example, when $\Omega_- = 1.99$, the frequencies and amplitudes of the forward waves are $(|\Omega_{F,1}| = 2.91,\ |F_1| = 0.54\text{A}_0)$ and $(|\Omega_{F,2}| = 0.68,\ |F_2| = 0.38\text{A}_0)$, those of the backward waves are $(|\Omega_{B,1}| = 2.91, |B_1| = 0.10\text{A}_0)$ and $\left(|\Omega_{B,2}| = 0.68, |B_2| = 0.18\text{A}_0\right)$. The discrepancies observed in Fig. 3 (b) are mainly attributed to the spatial Fourier transform procedure applied to the FDTD results.

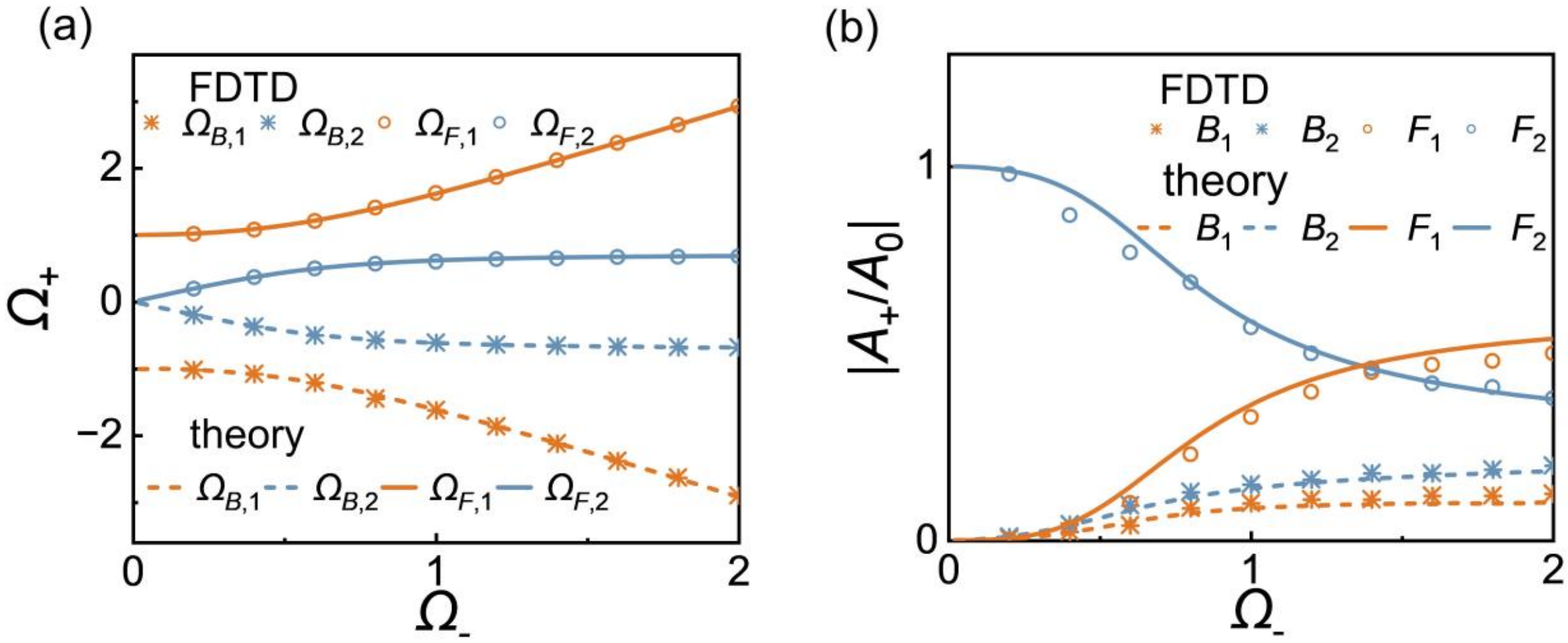

Fig. 3. Frequencies (a) and amplitudes (b) of the scattered waves versus incident wave's frequency.

3.2 *Another method for calculating the amplitudes of the temporally scattered waves*

Although Eqs. (9)-(11) enable calculation of the scattered-wave amplitudes, they do not explicitly indicate the factors that determine their magnitudes. To address this limitation, we derive analytical expressions for the scattered-wave amplitudes using a modal and impedance-based analysis across the temporal interface.

*3.2.1 Modal and impedance-based analysis*

First, to perform modal analysis, we assume a displacement solution of the form $[u_n, v_n^1]^{\mathrm{T}} = [u, v^1]^{\mathrm{T}} e^{i\Omega\tau - iqna}$, Eqs. (2a)-(2c) can be rewritten in the matrix form for $\tau > \tau_0$:

$$\left(-\Omega^2\mathbf{M} + \mathbf{K}\right)\boldsymbol{\varphi} = 0 \tag{12}$$

in which, $\boldsymbol{\varphi} = [u, v^1]^{\mathrm{T}}$ is the eigenvector. $\mathbf{M}$ and $\mathbf{K}$ are the mass matrix and stiffness matrix, respectively, with their expressions given by:

$$\mathbf{M} = \begin{bmatrix} 1 & 0 \\ 0 & \delta_+ \end{bmatrix}, \mathbf{K} = \begin{bmatrix} -\left(1+\dfrac{\kappa}{4\gamma^2}\right)\left(-2+e^{-iqa}+e^{iqa}\right) & \dfrac{\kappa}{2\gamma}\left(-1+e^{iqa}\right) \\ -\dfrac{\kappa}{2\gamma}\left(1-e^{-iqa}\right) & \kappa \end{bmatrix} \tag{13}$$

The value of $q$ is determined by the dispersion relation before the temporal interface: $\left(-2 + e^{iqa} + e^{-iqa}\right) = \Omega_-^2$. By solving Eq. (12) we can obtain four eigenvalues $\Omega_{F,1}$, $\Omega_{F,2}$, $\Omega_{B,1}$, $\Omega_{B,2}$ (which are equal to the roots in Eq. (5)) and four corresponding eigenvectors $\boldsymbol{\varphi}_{F,1}$, $\boldsymbol{\varphi}_{F,2}$, $\boldsymbol{\varphi}_{B,1}$, $\boldsymbol{\varphi}_{B,2}$. For the undamped case, the eigenvectors satisfy the following relationship:

$$\boldsymbol{\varphi}_{F,1} = \boldsymbol{\varphi}_{B,1} \tag{14a}$$

$$\boldsymbol{\varphi}_{F,2} = \boldsymbol{\varphi}_{B,2} \tag{14b}$$

$$\boldsymbol{\varphi}_i^{\dagger}\mathbf{M}\boldsymbol{\varphi}_j = 0, i \neq j \tag{14c}$$

$$\boldsymbol{\varphi}_i^{\dagger}\mathbf{M}\boldsymbol{\varphi}_j = M_{pi}, i = j \tag{14d}$$

here, $i, j = 1,2$. $M_{pi}$ denotes the $i$th principal mass. The symbol $\dagger$ means conjugate transpose.

In the following analysis, we denote the two orthogonal modes in the metamaterial after the interface as $\boldsymbol{\varphi}_1 = \boldsymbol{\varphi}_{F,1} = \boldsymbol{\varphi}_{B,1}$ and $\boldsymbol{\varphi}_2 = \boldsymbol{\varphi}_{F,2} = \boldsymbol{\varphi}_{B,2}$. We also define the wave mode in the chain before the temporal interface as $\boldsymbol{\varphi}_0$. For the convenience of analysis, the above wave modes $\boldsymbol{\varphi}_{m=0,1,2}$ are normalized using the following convention:

(i) $\|\boldsymbol{\varphi}_m\| = \sqrt{|u|^2 + |v^1|^2} = 1, m = 0,1,2$

(ii) $v^1 \in \Re,\ v^1 \geq 0$

Therefore, the following expressions can be derived based on the continuity of displacement (Eqs. (6a) and (7a)), with details provided in Appendix B.

$$\frac{F_1 + B_1}{A_0} = \frac{\left(\boldsymbol{\varphi}_1^{\dagger}\mathbf{M}\boldsymbol{\varphi}_0\right)}{M_{p1}}\left(\frac{\varphi_{11}}{\varphi_{01}}\right) \tag{15a}$$

$$\frac{F_2 + B_2}{A_0} = \frac{\left(\boldsymbol{\varphi}_2^{\dagger}\mathbf{M}\boldsymbol{\varphi}_0\right)}{M_{p2}}\left(\frac{\varphi_{21}}{\varphi_{01}}\right) \tag{15b}$$

in which, $\varphi_{01}$, $\varphi_{11}$ and $\varphi_{21}$ represents the first component in $\boldsymbol{\varphi}_0$, $\boldsymbol{\varphi}_1$ and $\boldsymbol{\varphi}_2$, respectively.

Moreover, the relationship between $F_{m(m=1,2)}, B_m$ and the impedances can be derived from Eqs. (6b) and (7b):

$$A_0 Z_0 = Z_1\left(F_1 - B_1\right) + Z_2\left(F_2 - B_2\right) \tag{16a}$$

$$Z_0^3 A_0\left(1 + \frac{\kappa}{4\gamma^2}\right) = \left(Z_1^3\left(F_1 - B_1\right) + Z_2^3\left(F_2 - B_2\right)\right) \tag{16b}$$

here, $Z_0$ and $Z_{m(m=1,2)}$ denote the impedances of the metamaterial before and after the interface at frequency $\Omega_m (= \Omega_{F,m} = -\Omega_{B,m})$, respectively. The detailed expressions of these impedances are provided in Appendix C. Note that Eqs. (16a) and (16b) are derived from Eqs. (6b) and (7b), they must be modified for other activation protocols of resonators.

### *3.2.2 Analytical expressions for amplitudes of the scattered waves*

The analytical expressions for the amplitudes of the temporally scattered waves can be derived from Eqs. (15a)-(16b):

$$\left|\frac{F_1}{A_0}\right| = \frac{1}{2}\left|\Pi\left(\boldsymbol{\varphi}_0, \boldsymbol{\varphi}_1\right) + \frac{Z_0}{Z_1}\frac{Z_2^2 - Z_-^2}{Z_2^2 - Z_1^2}\right| \tag{17a}$$

$$\left|\frac{B_1}{A_0}\right| = \frac{1}{2}\left|\Pi\left(\boldsymbol{\varphi}_0, \boldsymbol{\varphi}_1\right) - \frac{Z_0}{Z_1}\frac{Z_2^2 - Z_-^2}{Z_2^2 - Z_1^2}\right| \tag{17b}$$

$$\left|\frac{F_2}{A_0}\right| = \frac{1}{2}\left|\Pi(\boldsymbol{\varphi}_0, \boldsymbol{\varphi}_2) + \frac{Z_0}{Z_2}\frac{Z_1^2 - Z_-^2}{Z_1^2 - Z_2^2}\right| \tag{17c}$$

$$\left|\frac{B_2}{A_0}\right| = \frac{1}{2}\left|\Pi(\boldsymbol{\varphi}_0, \boldsymbol{\varphi}_2) - \frac{Z_0}{Z_2}\frac{Z_1^2 - Z_-^2}{Z_1^2 - Z_2^2}\right| \tag{17d}$$

Here, we define $\Pi(\boldsymbol{\varphi}_0, \boldsymbol{\varphi}_m) = \frac{\left(\boldsymbol{\varphi}_m^{\dagger}\mathbf{M}\boldsymbol{\varphi}_0\right)}{M_{pm}}\left(\frac{\varphi_{m1}}{\varphi_{01}}\right)$ as the weighted modal correlation coefficients between modes $\boldsymbol{\varphi}_0$ and $\boldsymbol{\varphi}_{m=1,2}$, and $Z_- = Z_0\sqrt{\frac{\kappa}{4\gamma^2} + 1}$.

Eqs. (17a) to (17d) reveal that the amplitudes of temporally scattered waves are determined by the weighted modal correlation coefficients and the impedances across the interface. This relationship provides a mechanism for controlling the scattering behavior by appropriately tailoring the modal structures and impedances before and after the temporal interface. It should be noted that, Eqs. (17a) to (17d) can also be used to predict the amplitudes of temporally scattered waves in small damping cases. But they no longer work when $\zeta$ becomes too large, because Eqs. (14a)-(14d) cease to be satisfied.

## 4. Temporal evanescent wave

To determine how damping reshapes the temporal-interface response, Scenario *II* is examined in this section. When damping in the resonators is considered ($\zeta \neq 0$), frequency splitting can still be observed. Notably, when the damping coefficient exceeds a critical threshold, exceptional points (EPs) emerge, accompanied by the appearance of temporal evanescent waves characterized by purely imaginary frequencies and real wavenumbers.

### *4.1 Influences of damping on the reflected and refracted waves*

When damping is introduced, frequencies and amplitudes of the scattered waves can also be evaluated using the theory established in Section 3.1 with several modifications. Eq. (4) is generalized to be:

$$\frac{\Omega_+^2}{\Omega_0^2} = \frac{\Omega_-^2}{\Omega_0^2} + \frac{\kappa}{4\gamma^2}\left(1 - \frac{1}{-\frac{\Omega_+^2}{\Omega_0^2} + 1 + 2i\zeta\frac{\Omega_+}{\Omega_0}}\right)\frac{\Omega_-^2}{\Omega_0^2} \tag{18}$$

Just like in Eq. (5), $\Omega_+$ and $\Omega_-$ can be normalized by dividing by $\Omega_0$. Therefore, we also set $\Omega_0 = 1$ during the analysis in this section.

Eq. (18) indicates that there are still four solutions for $\zeta \neq 0$: $\Omega_{F,1}$, $\Omega_{B,1}$ and $\Omega_{F,2}$, $\Omega_{B,2}$. But these frequencies must be complex to satisfy the complex dispersion relation in the damped metamaterial. To calculate the amplitudes of these scattered waves, the matrix $\mathbf{M}_s$ in Eq. (11) should be modified to be:

$$\mathbf{M}_s = \begin{bmatrix} 1 & 1 & 1 & 1 \\ \Omega_{F,1} & \Omega_{F,2} & \Omega_{B,1} & \Omega_{B,2} \\ \dfrac{\Omega_0^2}{-\left(\Omega_{F,1}\right)^2 + \Omega_0^2 + 2i\zeta\Omega_0\Omega_{F,1}} & \cdots & \cdots & \dfrac{\Omega_0^2}{-\left(\Omega_{B,2}\right)^2 + \Omega_0^2 + 2i\zeta\Omega_0\Omega_{B,2}} \\ \dfrac{\Omega_{F,1}\Omega_0^2}{-\left(\Omega_{F,1}\right)^2 + \Omega_0^2 + 2i\zeta\Omega_0\Omega_{F,1}} & \cdots & \cdots & \dfrac{\Omega_{B,2}\Omega_0^2}{-\left(\Omega_{B,2}\right)^2 + \Omega_0^2 + 2i\zeta\Omega_0\Omega_{B,2}} \end{bmatrix} \tag{19}$$

Fig. 4 presents the changes in frequencies and amplitudes of the scattered waves as a function of damping in the resonator when $\Omega_- = 1$. It can be seen that when $\zeta$ is increased from zero, the imaginary parts of $\Omega_{F,2}$ and $\Omega_{B,2}$ gradually increase, while their real parts gradually approach zero. This trend continues until reaching a critical point (highlighted by $\zeta_C$ in the figures). At this point, $\Omega_{F,2}$ and $\Omega_{B,2}$ collapses into a same purely imaginary frequency $\Omega_c = 0.67i$. After the critical point $\zeta_c$, $\Omega_{F,2}$ and $\Omega_{B,2}$ are again spilt into two different imaginary values.

Amplitudes of the waves with purely imaginary frequencies are nonzero (see Fig. 4(c)), which means they can be observed after the interface. Since these waves have imaginary frequencies but real wave numbers, their group velocities will also be imaginary. As a result, we can predict that these waves cannot propagate in space and their amplitudes will exponentially decrease in the time domain. By analogy with spatial evanescent waves, we refer to them as temporal evanescent waves in this study.

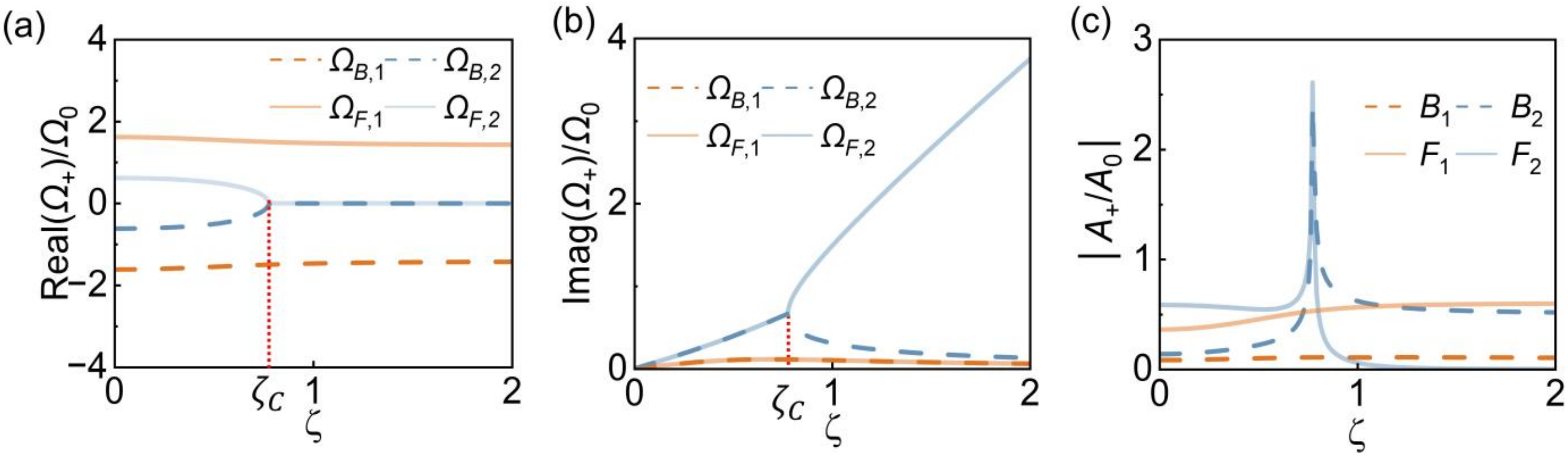

Fig. 4. Frequencies (imaginary (a) and real (b) parts) and amplitudes (c) of scattered waves versus $\zeta$. The vertical dotted line in (a) and (b) marks the position of a critical damping $\zeta_C$.

To verify the existence of temporal evanescent waves, a FDTD simulation is carried out, the results are shown in Fig. 5. In the simulation, the interface is created by abruptly changing $\delta$ from zero to $\delta_+ = 1$, and $\zeta$ from zero to $\zeta_+ = 5$. A tone-burst wave packet with central frequency at $\Omega_- = 1$ is propagating in the chain before the interface, as illustrated in Fig. 5(a). It is scattered by the interface at $\tau = \tau_0$. After the interface, part of the energy remains confined at the position corresponding to $\tau = \tau_0$, maintaining its original waveform without propagating, and gradually attenuates over time. To further examine this localized mode, in Fig. 5(b), the time evolution of the normalized displacement is illustrated for the mass located at $x_c = 29.6\lambda$, corresponding to the wave packet center at $\tau = \tau_0$, as denoted by the vertical dash-dot line in Fig. 5(a). In the initial period following $\tau_0$, the vibration at this point is a superposition of all the scattered waves. As the forward and backward propagating waves gradually separate, only the temporal evanescent wave with purely imaginary frequency is remained, manifesting as non-oscillatory exponential decay in time domain. The dashed line represents the theoretical prediction of the normalized displacement for the temporal evanescent component, calculated using Eqs. (18) and (19). The FDTD simulation results show excellent agreement with the theoretical curve, confirming the presence of a pure temporal evanescent wave.

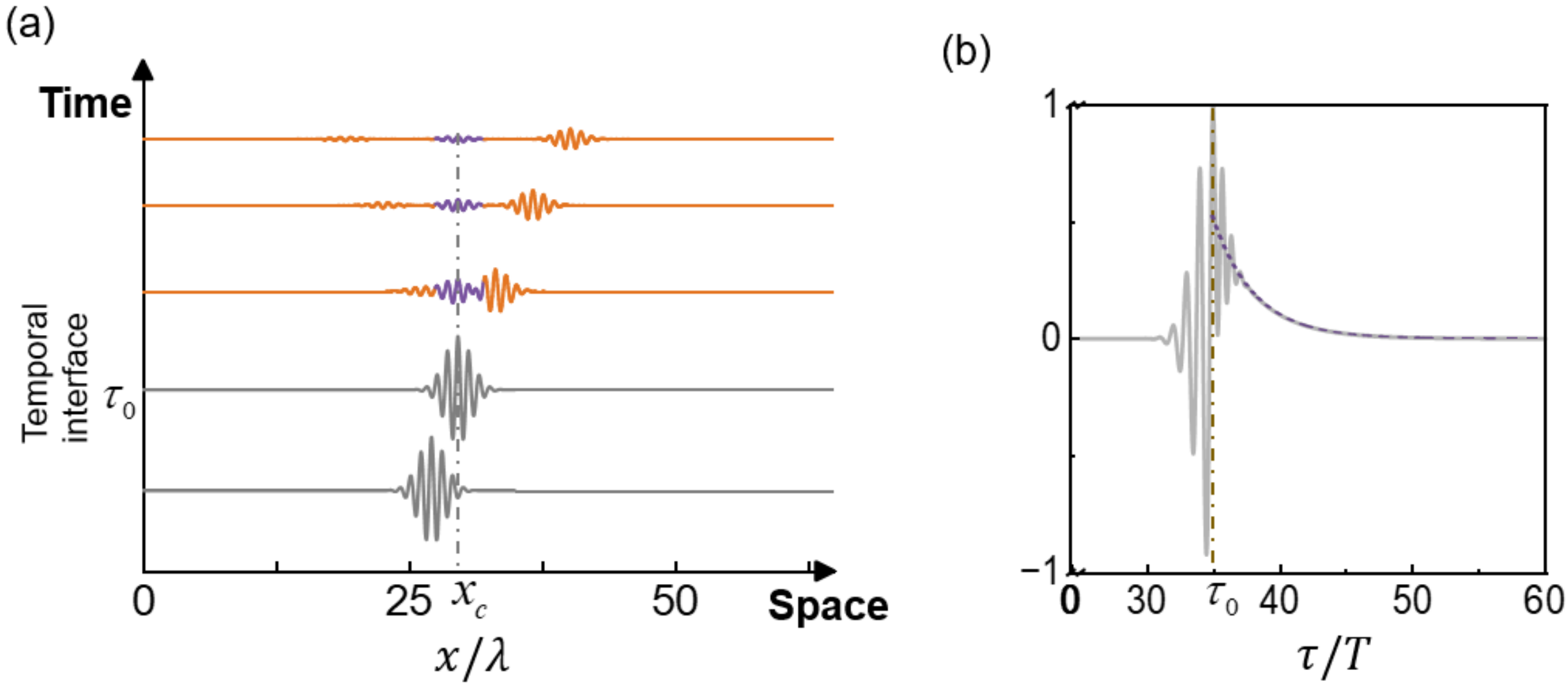

Fig. 5. (a) Propagation of wave packet before and after the interface ($\tau = \tau_0$) simulated using the FDTD method. The dash-dot line marks the position of the mass at $x_c = 29.6\lambda$. (b) Time evolution of the normalized displacement of the mass located at $x_c = 29.6\lambda$, the displacement is normalized by dividing the amplitude of the incident wave, $T$ is the period of the incident wave at $\Omega_- = 1$, the solid line represents the result obtained using the FDTD method, the dashed line represents the theoretical result.

It should be emphasized that the temporal evanescent wave reported here is fundamentally different from the decaying response of an overdamped vibration system, although both are associated with imaginary eigenvalues. The temporal evanescent wave reported here originates from the material properties across the temporal interface, as will be discussed in detail in Section 4.3, rather than from the structural feature of a finite vibration system.

*4.2 Conditions for the emergence of temporal evanescent waves*

We first demonstrate that the critical point highlighted in Fig. 4 is an exceptional point (EP), caused by the coalescence of two resonator-dominated modes. Based on this understanding, we provide the parameter conditions under which the temporal evanescent wave emerges.

When $\zeta \neq 0$, Eq. (12) can be rewritten as:

$$\left(-\Omega^2\mathbf{M} + i\Omega\mathbf{C} + \mathbf{K}\right)\boldsymbol{\varphi} = 0 \tag{20}$$

where $\mathbf{M}$ and $\mathbf{K}$ are defined in Eq. (13), and $\mathbf{C}$ denotes the dissipation matrix given by:

$$\mathbf{C} = \begin{bmatrix} 0 & 0 \\ 0 & 2\zeta\delta_+\Omega_0^+ \end{bmatrix} \tag{21}$$

By solving the above equations, there are also four eigenvalues $\Omega_{F,1}$, $\Omega_{F,2}$, $\Omega_{B,1}$, $\Omega_{B,2}$ and four corresponding eigenvectors $\boldsymbol{\varphi}_{F,1}$, $\boldsymbol{\varphi}_{F,2}$, $\boldsymbol{\varphi}_{B,1}$, $\boldsymbol{\varphi}_{B,2}$. However, different from the undamped case in Section 3.2, the non-zero $\zeta$ leads to a non-Hermitian system where the parallel or orthogonal relationships between the eigenvectors in Eqs. (14a)- (14b) no longer hold in most instances:

$$\boldsymbol{\varphi}_{F,1} \neq \boldsymbol{\varphi}_{B,1} \tag{22a}$$

$$\boldsymbol{\varphi}_{F,2} \neq \boldsymbol{\varphi}_{B,2} \tag{22b}$$

Only at $\zeta_c$, the two resonator-dominated modes $\boldsymbol{\varphi}_{B,2}$ and $\boldsymbol{\varphi}_{F,2}$ coalesce again as the damping increases (see Fig. 6(a)), indicating that at this critical point not only the two eigenvalues (see Fig. 4(a) and (b)) are coalesced but also the corresponding two eigenvectors. Therefore, this critical point is an EP. To demonstrate the emergence of EP more clearly, we select two parameter points located before ($\zeta = 0.1$) and after ($\zeta = 1$) the EP (namely, point A and C in Fig. 6). At $\zeta = 0.1$, $\boldsymbol{\varphi}_{B,2} = [0.24 + 0.47i, 0.85]$ and $\boldsymbol{\varphi}_{F,2} = [0.29 + 0.44i, 0.85]$ exhibit identical amplitudes but distinct phase. Whereas at $\zeta = 1$, $\boldsymbol{\varphi}_{B,2} = [0.22 + 0.37i, 0.90]$ and $\boldsymbol{\varphi}_{F,2} = [0.12 + 0.20i, 0.97]$ share identical phases but different amplitudes. Crucially, only precisely at the EP, both eigenvalues and eigenvectors coalesce into a single degenerate state simultaneously with $\boldsymbol{\varphi}_{B,2} = \boldsymbol{\varphi}_{F,2} = [0.19 + 0.33i, 0.93]$.

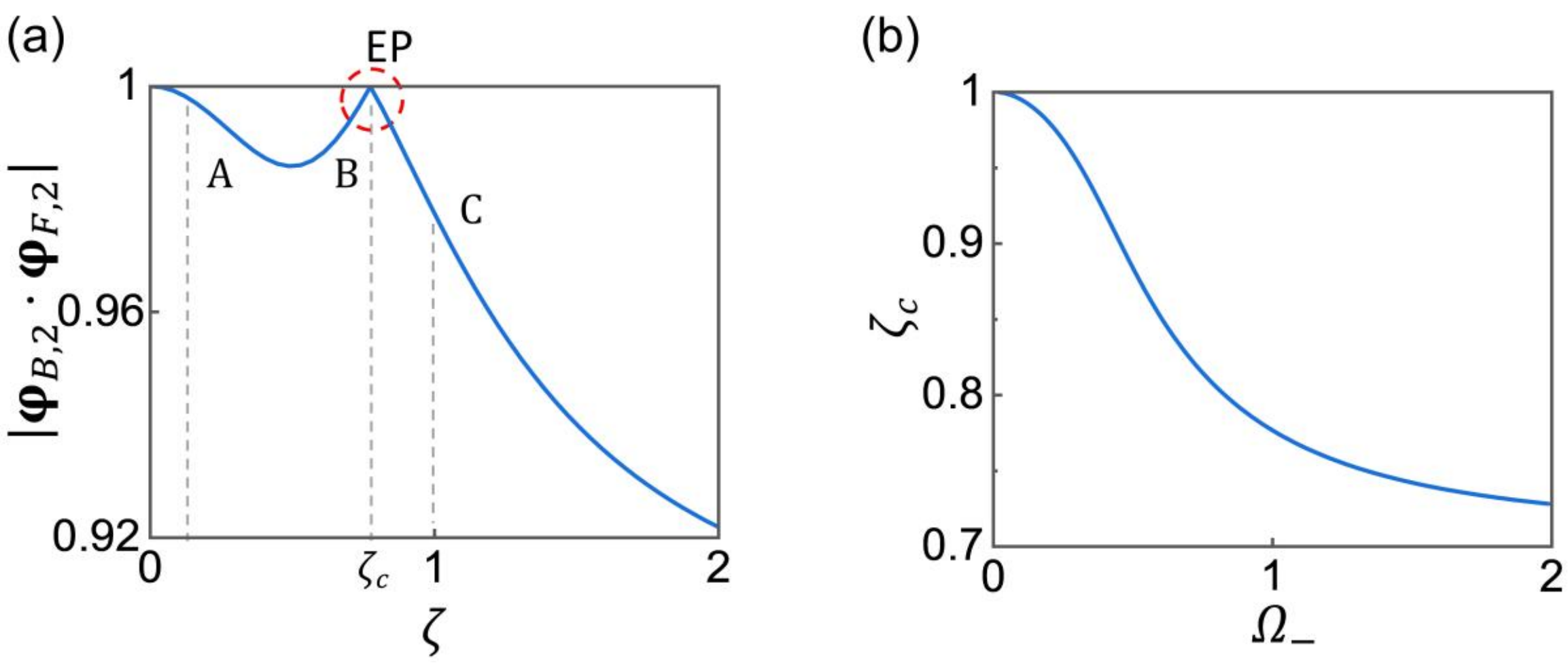

Fig. 6 (a) The modal correlation coefficient between $\boldsymbol{\varphi}_{B,2}$ and $\boldsymbol{\varphi}_{F,2}$ versus $\zeta$, the three points A, B, C correspond to the parameters $\zeta = 0.1$, $\zeta = \zeta_c = 0.78$ and $\zeta = 1$. (b) The critical damping $\zeta_C$ versus the incident frequency $\Omega_-$.

Note that this non-Hermitian degeneracy originates from dissipation, it can be readily verified that the dissipation matrix **C** is non-proportional, which is consistent with the conclusion reported by (Phani, 2022).

Since the appearance of temporal evanescent waves is indicated by an EP, it is desirable to directly predict the location of EP (i.e., the value of $\zeta_C$) without solving Eq. (18) within a wider parameter space.

By setting the determinant of the matrix in Eq. (20) to be zero leads to the following characteristic equation:

$$\left(-\Omega^2-\left(1+\frac{\kappa}{4\gamma^2}\right)\left(-2+e^{iqa}+e^{-iqa}\right)\right)\left(-\Omega^2+2i\zeta\Omega_0\Omega+\frac{\kappa}{\delta}\right)+\frac{\kappa\Omega_0^2}{4\gamma^2}\left(-2+e^{iqa}+e^{-iqa}\right)=0 \tag{23}$$

If Eq. (23) has two repeated roots, it can be written in the following form:

$$\left(\Omega-\Omega_c\right)^2\left(\Omega-\Omega_m\right)\left(\Omega-\Omega_n\right)=0 \tag{24}$$

where $\Omega_c$ is the degenerated eigenvalue, and $\Omega_m$, $\Omega_n$ donate the other two eigenvalues of the characteristic equation. By matching the coefficients of corresponding terms in Eq. (23) and Eq. (24), we derive the conditions under which repeated eigenvalues occur:

$$2\Omega_c\left(2i\zeta\Omega_0-2\Omega_c\right)+\frac{\left(\Omega_-^2\Omega_0^2\right)}{\left(\Omega_c\right)^2}+\left(\Omega_c\right)^2=-\left(\Omega_0^2+\left(1+\frac{\kappa}{4\gamma^2}\right)\Omega_-^2\right) \tag{25a}$$

$$\frac{\left(\Omega_-^2\Omega_0^2\right)}{\left(\Omega_c\right)}+\left(\Omega_c\right)^2\left(i\zeta\Omega_0-\Omega_c\right)=-\left(1+\frac{\kappa}{4\gamma^2}\right)\Omega_-^2 i\zeta\Omega_0 \tag{25b}$$

Once the incident frequency $\Omega_-$ and resonance frequency $\Omega_0$ are given, solving Eqs. (25a) and (25b) yields six pairs of solutions $(\Omega_c,\zeta)$. The particular solution $\zeta_C$ corresponding to a purely imaginary frequency $\Omega_c$ represents the critical threshold above which temporal evanescent modes begin to emerge. When material parameters across an interface are given, $\zeta_C$ depends on the incident frequency $\Omega_-$, as quantitatively demonstrated in Fig. 6(b).

### *4.3 Mechanism of temporal evanescent modes*

To elucidate the mechanism of temporal evanescent waves, we use an equivalent model with effective stiffness (Fig. 7(a)) to describe the locally resonant metamaterial in Fig. 1(a). By assuming time-harmonic quantities and eliminating $v_n^0, v_n^1$ in Eqs. (2a)-(2c), the dimensionless effective stiffness can be deduced as the following expression:

$$\kappa_{\text{eff}} = 1 + \frac{\kappa}{4\gamma^2}\left(1 - \frac{\Omega_0^2}{-\Omega^2 + \Omega_0^2 + 2i\zeta\Omega\Omega_0}\right) \tag{26}$$

We examine the influences of $\zeta$ on effective stiffness of the metamaterial at $\Omega_{F,2}$ and $\Omega_{B,2}$ in Fig. 7(b) and (c). Before the EP, $\kappa_{F,2} = \kappa_{\text{eff}}\ (\Omega_{F,2})$ and $\kappa_{B,2} = \kappa_{\text{eff}}\ (\Omega_{B,2})$ form a pair of complex conjugates (i.e., $\kappa_{B,2} = (\kappa_{F,2})^*$), with the magnitude of their imaginary parts first increasing and then decreasing. Notably, at and after the EP, $\kappa_{F,2}$ and $\kappa_{B,2}$ become negative real numbers, suggesting that the temporal evanescent modes originate from negative effective stiffness.

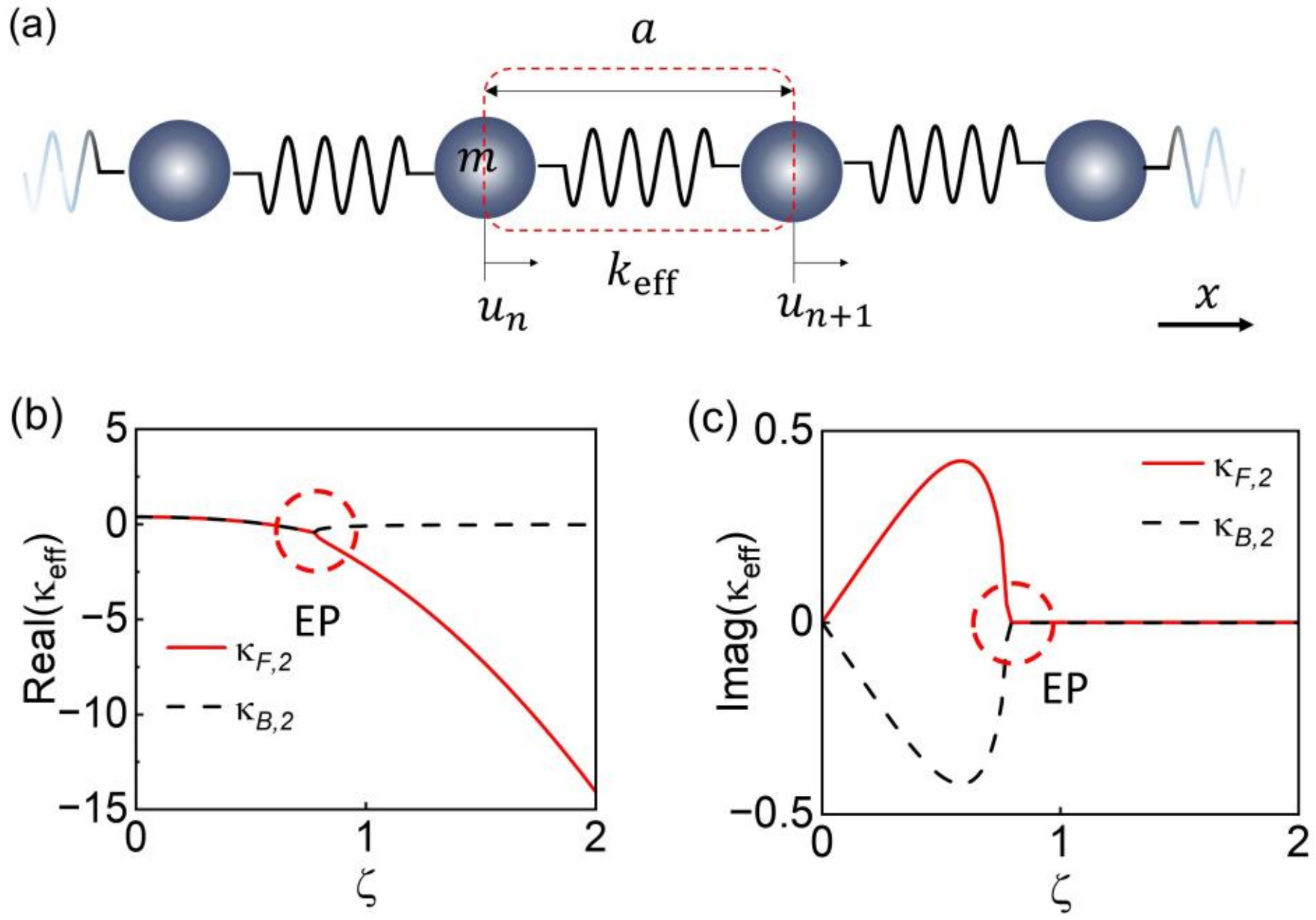

Fig. 7. (a) The effective model of the metamaterial in Fig. 1(a). Imaginary (b) and real (c) parts of the effective stiffness ($\kappa_{F,2} = \kappa_{\text{eff}}\ (\Omega_{F,2})$, $\kappa_{B,2} = \kappa_{\text{eff}}\ (\Omega_{B,2})$) versus $\zeta$.

It must be emphasized that, at and after the EP, $\Omega_{F,2}$ and $\Omega_{B,2}$ are all imaginary (see Fig. 4), which means the temporal evanescent waves are caused by negative effective material parameters defined on the imaginary frequency axis, totally different from the existing ones defined on the real frequency axis (Liu et al., 2000; Fang et al., 2006; Zhou et al., 2012). The distinctions between these two definitions are clarified hereinafter.

First, we revisit the traditional definition of negative effective material parameters by studying the unit cell in Fig. 8(a). Eq. (26), expressed in dimensional form, can be used to calculate the effective stiffness by neglecting damping: $k_{\text{eff}} = k_p + \frac{k_1}{4\tan^2\theta}(1 - \frac{\omega_0^2}{-\omega^2+\omega_0^2})$, where $\omega$ may take complex values. The variation of the effective stiffness with respect to real valued frequency Re(ω) is shown in Fig. 8 (b), the parameters are chosen as $k_p = k_1 = 1, m_1 = 1, \tan\theta = 1/2$, in consistent with the nondimensional parameters used in Fig. 5 except the damping. To understand the mechanism behind the negative effective stiffness in this case, alternatively, we can also define the effective stiffness using $k_{\text{eff}} = \frac{F_e}{2|x|}$, as shown in Fig. 8(a), $x$ is the displacement of mass m, $F_e$ is the net force exerted on mass m by the rigid rod and the spring with stiffness $k_p$, it can be calculated through $F_e = F_p + F_T \cos\theta$, with $F_T$ being the force in the rigid rod, $F_p = 2k_p x$ being the force of spring $k_p$. Actually, $F_T = \frac{F_m}{2sin\theta}$, where $F_m = -m_1\omega^2 v_n^1$ is the inertial force of $m_1$. Therefore, we can derive $F_e = F_p + \frac{F_m}{2tan\theta}$. The variation of these forces with respect to real-valued frequency $\text{Re}(\omega)$ is shown in Fig. 8(c). At most frequencies, $F_e$ has positive value, therefore the effective stiffness is positive. When $\text{Re}(\omega) \to \omega_0 = \sqrt{k_1/m_1}$, the resonant effect significantly amplifies the acceleration of $m_1$, producing a substantial inertial force $F_m$. This large $F_m$ causes $F_e$ to become negative, leading to a negative effective stiffness. Therefore, negative effective stiffness defined on the real frequency axis are caused by the large inertial force of the mass in the resonator.

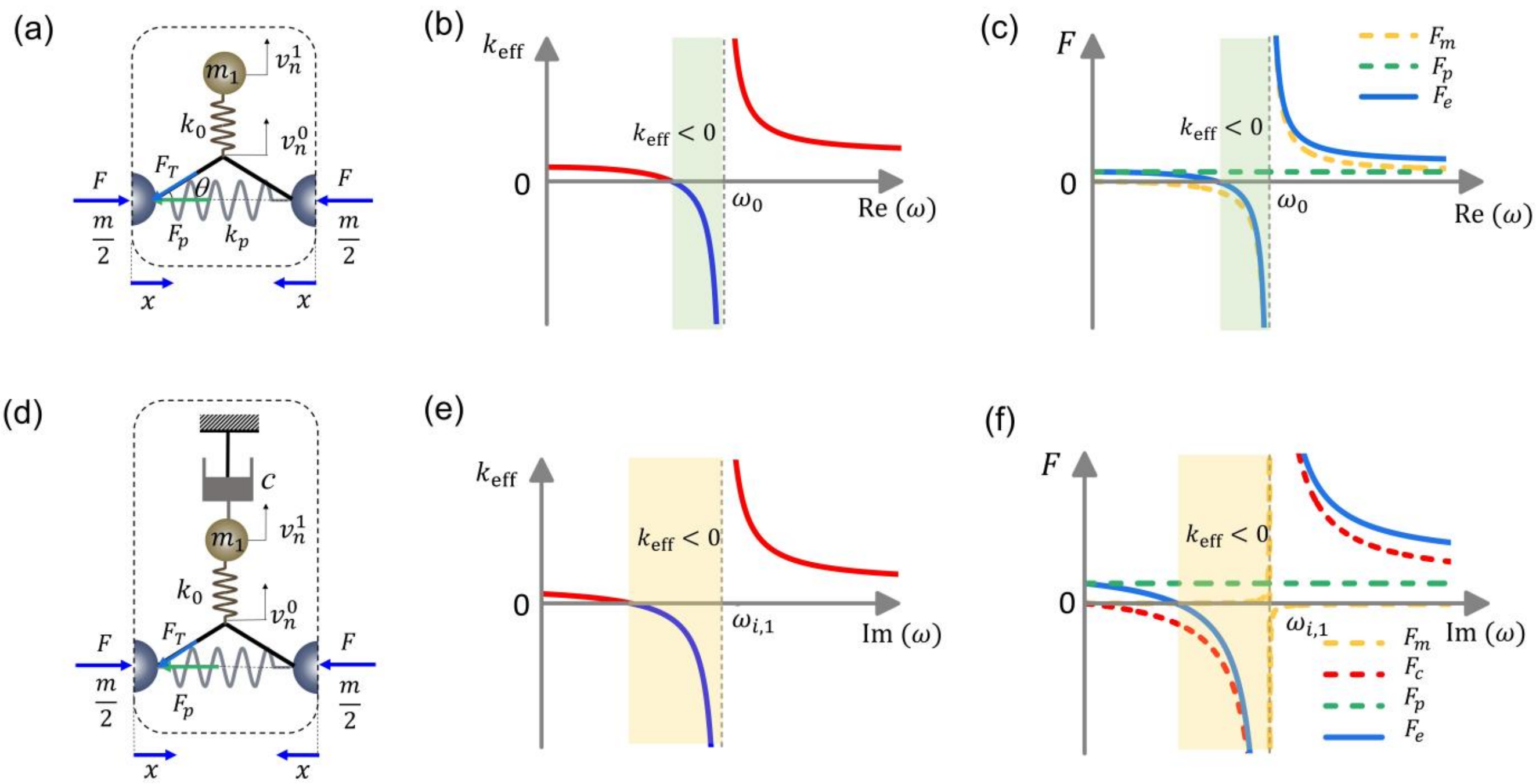

Fig. 8. (a) A unit cell without damping, $F_p$ is the force in spring $k_p$, $F_T$ is the force in the rigid rod. (b) Variation of the effective stiffneness of the unit cell in (a) with respect to real-valued frequency. (c) Variation of the force $F_m$, $F_p$ and $F_e = F_p + \frac{F_m}{2tan\theta}$ with respect to real-valued frequency, $F_m$ is the inertial force of $m_1$. The vertical dashed line represents the frequency $\omega_0$ in (b) and (c). (d) A unit cell with damping. (e) Variation of the effective stiffneness of the unit cell in (d) with respect to imaginary-valued frequency. (f) Variation of the force $F_m$, $F_p$, $F_c$ and $F_e = F_p + \frac{F_m+F_c}{2tan\theta}$ with respect to imaginary -valued frequency, $F_c$ is the damping force acting on the mass $m_1$. The vertical dashed line represents the frequency $\omega_{i,1}$ in (e) and (f).

Now, we consider the effective stiffness of the unit cell in Fig. 8(d), a damper $c$ is included in it compared with the unit cell in Fig. 8(a). As discussed in Appendix D, the number of frequency bands exhibiting negative effective stiffness depends on the value of damping ratio. When $\zeta > 1$, two frequency bands satisfying $k_{\text{eff}} < 0$ can be observed, the upper limit $\omega_{i1}$ for the first band and the lower limit $\omega_{i2}$ for the second one can be obtained by solving $-\omega^2 + \omega_0^2 + 2i\zeta\omega_0\omega = 0$. When $\zeta_c \leq \zeta \leq 1$, only one band with $k_{\text{eff}} < 0$ exists, whose boundaries are determined by solving $k_{\text{eff}} = 0$. In either case, the underlying mechanism for negative effective stiffness formation remains identical. Therefore, the first frequency band in Fig. 8(e) for $\zeta = 5$ is taken as an example, with other parameters consistent with the undamped unit cell. Similar to the undamped case, $F_e$ can be derived as $F_e = F_p + \frac{F_m+F_c}{2tan\theta}$, where $F_c = ic\omega v_n^1$ denotes the damping force acting on the mass $m_1$. Variation of the forces with respect to imaginary frequency is shown in Fig. 8(f). $F_c$ becomes significantly amplified as $\text{Im}(\omega) \rightarrow \omega_{i1}$, resulting in a negative $F_e$ and $k_{\text{eff}}$, Therefore, negative material parameters defined on imaginary frequency axis are mainly due to the large damping force in the damped resonator.

In the end, Fig. 9 compares the mechanisms, fundamental concepts, and characteristics of temporal and spatial evanescent waves. As the temporal analogs of spatial evanescent waves, temporal evanescent waves also originate from an interface between positive and negative material parameters and exhibit localization near the interface, as shown in Fig. 9. However, these two types of evanescent mode have significant differences. A temporal evanescent wave is characterized by an imaginary frequency and a real wavenumber, while its spatial counterpart exhibits a real frequency and imaginary wavenumber. A temporal evanescent wave may exist throughout the spatial domain but is confined to a limited time interval, in the contrary, a spatial evanescent wave is confined to a limited space but has no attenuation in the time domain. Causality imposes additional constraints, requiring that temporal evanescent waves only exist after the temporal interface, whereas spatial evanescent waves extend on both sides of the spatial interface. These differences make temporal evanescent modes unique and may be exploited to achieve new unprecedented wave-based functionalities.

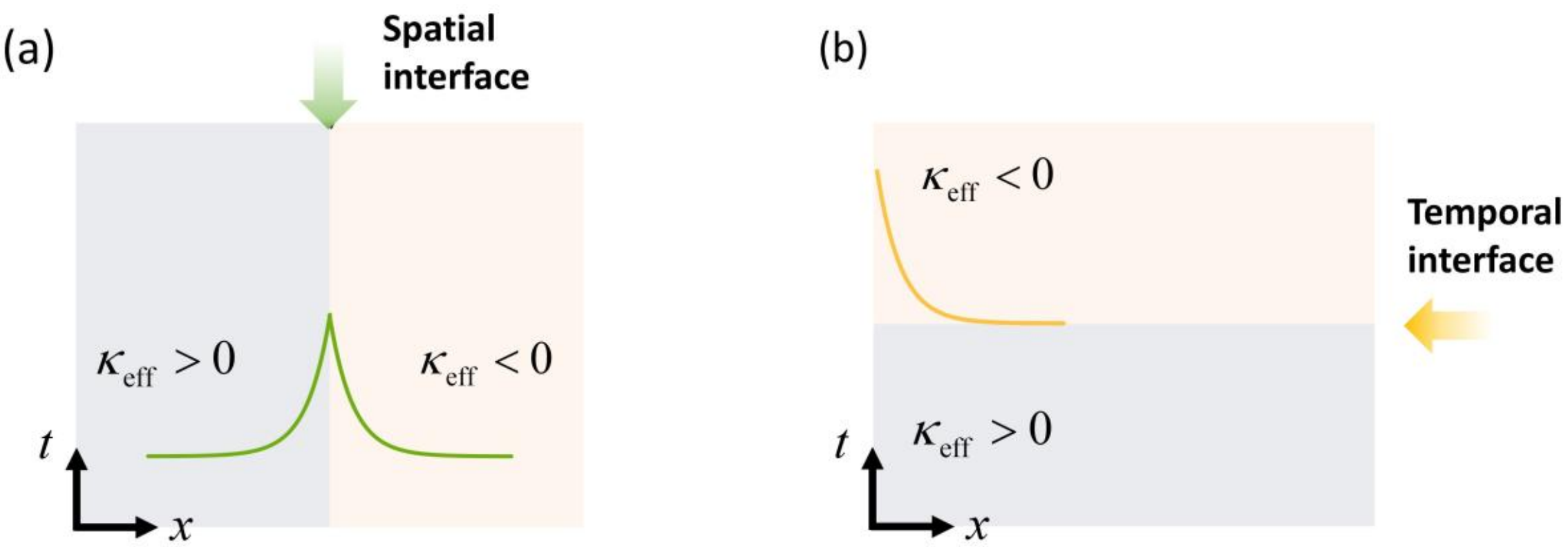

Fig. 9. (a) Spatial evanescent mode at a spatial interface between two semi-infinite media with two oppositely-signed effective stiffness. (b) Temporal evanescent mode at a temporal interface between two semi-infinite media with two oppositely-signed effective stiffness.

## 5. Conclusion

This work presents a theoretical investigation of wave scattering at temporal interfaces formed by the abrupt activation of local resonators in a one-dimensional elastic metamaterial. The sudden onset of resonance induces a temporal discontinuity, across which a monochromatic incident wave generates two distinct frequency components in both forward and backward directions, consistent with the second-order dispersion relation. By deriving analytical expressions, we demonstrate that the amplitudes of the scattered waves are governed by the weighted modal correlation coefficients and the impedances across the interface. These insights suggest a promising strategy for tailoring wave spectra through careful design of pre- and post-interface material parameters.

Additionally, we identify the emergence of temporal evanescent modes when damping in the resonators exceeds an EP, which is caused by the coalescence of two resonator-dominated modes. These temporal evanescent waves exhibit non-propagating behavior, maintaining their instantaneous waveform at the interface while decaying exponentially over time, representing a novel type of wave mode. Analytical expressions are derived to predict their existence. From a material perspective, these evanescent modes arise from a negative effective stiffness evaluated at imaginary frequencies, which is induced by the dominant damping forces within the microstructure.

The discovery of frequency-splitting phenomena and temporal evanescent modes expands the scope of time-varying mechanics and opens new pathways for wave-based control and signal processing. While this study focuses on a one-dimensional setting, the established theoretical framework and underlying physics is expected to extend to higher-dimensional systems and continuum elastic media.

**Acknowledgements**

We acknowledge the support from National Natural Science Foundation of China (grant numbers 12572100, 12532006).

**Appendix A. Wave scattering at a temporal interface obtained via resonator stiffness switching**

This appendix analyzes wave scattering at a temporal interface created by activating the resonator through an abrupt change in $k_1$ from zero to non-zero. Both key phenomena predicted by the proposed theoretical framework—frequency splitting and temporal evanescent waves—are observed in this case.

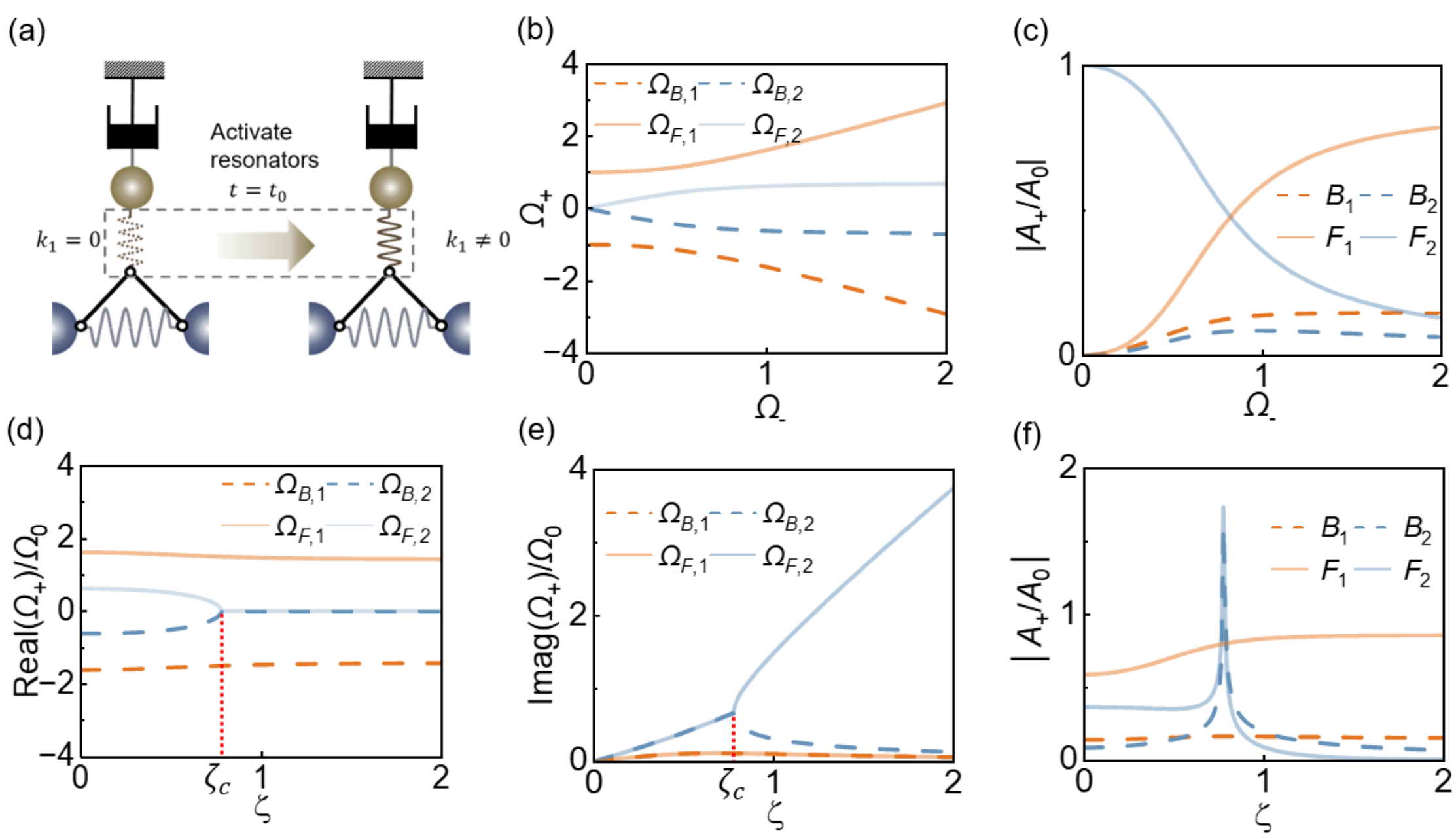

Fig.A1 (a) Schematic illustration of the abrupt changes of $k_1$ at $t_0$. (b)(c) Frequencies (b) and amplitudes (c) of the scattered waves versus incident frequency. (d-f) Frequencies (real (d) and imaginary (e) parts) and amplitudes (f) of scattered waves versus ζ. The vertical dotted line in (d) and (e) marks the position of a critical damping $\zeta_c$.

Fig. A1(a) illustrates how the temporal interface is introduced. For $t < t_0$, $k_1 = 0$, and the resonators are deactivated. At $t = t_0$ the resonators are suddenly activated by switching $k_1$ from zero to a finite value. Figs. A1(b)(c) and (d-f) illustrate the wave scattering results in the absence and presence of damping, respectively. Comparison with Fig. 3 and Fig. 4 indicates that both frequency splitting and temporal evanescent waves are still observed. This is because different resonator activation strategies modify only the TBCs, without affecting the dispersion relations before and after the interface.

**Appendix B. Details for modal analysis**

In this Appendix, we provide details of the analysis of wave modes before and after the temporal interface. Waves before and after the temporal interface can be expressed as the superposition of the defined modes:

$$\boldsymbol{\psi}_- = \bar{A}_0 \boldsymbol{\varphi}_0 e^{i\Omega_- \tau - iqna} \tag{B1}$$

$$\boldsymbol{\psi}_+ = \sum_{m=1}^{2} \bar{F}_m \boldsymbol{\varphi}_m e^{i\Omega_{F,m}\tau - iqna} + \sum_{m=1}^{2} \bar{B}_m \boldsymbol{\varphi}_m e^{i\Omega_{B,m}\tau - iqna} \tag{B2}$$

in which, $\overline{A_0}$ is the modal amplitude of the incident wave. $\overline{F_1}$, $\overline{F_2}$, $\overline{B_1}$, $\overline{B_2}$ are the modal amplitudes of waves after the temporal interface. Based on the continuity of displacement across the temporal boundary (i.e. $\boldsymbol{\psi}_- |_{\tau=\tau_0} = \boldsymbol{\psi}_+ |_{\tau=\tau_0}$), we can obtain the following relation:

$$\bar{A}_0 \boldsymbol{\varphi}_0 e^{i\Omega_- \tau_0} e^{-iqna} = \left( \bar{F}_1 \boldsymbol{\varphi}_1 e^{i\Omega_{F,1}\tau_0} + \bar{F}_2 \boldsymbol{\varphi}_2 e^{i\Omega_{F,2}\tau_0} + \bar{B}_1 \boldsymbol{\varphi}_1 e^{i\Omega_{B,1}\tau_0} + \bar{B}_2 \boldsymbol{\varphi}_2 e^{i\Omega_{B,2}\tau_0} \right) e^{-iqna} \tag{B3}$$

Without loss of generality, we set $\tau_0 = 0$, since any switching instant can be shifted to $\tau_0' = 0$ through the transformation $\tau' = \tau - \tau_0$. Multiplying the left side of Eq. (B3) by $\mathbf{M}$, followed by left-multiplication with $\boldsymbol{\varphi}_1^\dagger$ or $\boldsymbol{\varphi}_2^\dagger$, the following relationships can be obtained using Eqs. (14a)-(14d):

$$\bar{A}_0 \left( \boldsymbol{\varphi}_1^\dagger \mathbf{M} \boldsymbol{\varphi}_0 \right) = \left( \bar{F}_1 + \bar{B}_1 \right) M_{p1} \tag{B4}$$

$$\bar{A}_0 \left( \boldsymbol{\varphi}_2^\dagger \mathbf{M} \boldsymbol{\varphi}_0 \right) = \left( \bar{F}_2 + \bar{B}_2 \right) M_{p2} \tag{B5}$$

Furthermore, by comparing Eqs. (8a)(8b) and Eqs. (B1)(B2), we have:

$$\bar{A}_0 \varphi_{01} = A_0 \tag{B6}$$

$$\bar{F}_m \varphi_{m1} = F_m \tag{B7}$$

$$\bar{B}_m \varphi_{m1} = B_m \tag{B8}$$

here, $m = 1,2$. Substituting Eqs. (B6)- (B8) into Eqs. (B4) and (B5), one can directly obtains Eqs. (15a) and (15b) in Section. 3.

**Appendix C. Details for impedance-based analysis**

In this Appendix, the detailed derivation of Eqs. (16a) and (16b) is presented. The derivation is first performed from the physical model given by Eqs. (1a)-(1c). Based on these results, the corresponding dimensionless form, i.e., the impedance relations of the dimensionless model given by Eqs. (2a)-(2c), can be directly written as Eqs. (16a) and (16b).

Based on the continuity of the momentum of mass $m$ in the horizontal chain, we can obtain the following relation:

$$i\omega_{-} A_0 e^{i\omega_{-} t_0} e^{-iqna} = \left( i\omega_1 F_1 e^{i\omega_1 t_0} + i\omega_2 F_2 e^{i\omega_2 t_0} - i\omega_1 B_1 e^{-i\omega_1 t_0} - i\omega_2 B_2 e^{-i\omega_2 t_0} \right) e^{-iqna} \tag{C1}$$

where $\omega_{-}$ is the incident frequency, and $\omega_1 = \omega_{F,1} = -\omega_{B,1}$, $\omega_2 = \omega_{F,2} = -\omega_{B,2}$ represent the frequencies of the scattered waves. Assuming $t_0 = 0$ and using the conservation of wavenumber, the following expression can be obtained:

$$c_0 A_0 = c_1 (F_1 - B_1) + c_2 (F_2 - B_2) \tag{C2}$$

where $c_0$ and $c_j$ $(j = 1,2)$ are the phase velocities before and after the interface. Multiple both sides by $m$, Eq. (C2) can be rewritten as:

$$Z_0 A_0 = Z_1 (F_1 - B_1) + Z_2 (F_2 - B_2) \tag{C3}$$

where $Z_0 = \sqrt{mk_p}\cos\frac{qa}{2}$, $Z_j = \sqrt{m\left( k_p + \frac{k_1}{4\gamma^2}(1 - \frac{\omega_0^2}{-\omega_j^2 + \omega_0^2}) \right)}\cos\left(\frac{qa}{2}\right)$, the detailed derivation for impedances can be found in (Kim, 2023).

Based on the continuity of the momentum of mass $m_1$, we can also obtain the following relation:

$$0 = m_1 \left( i\omega_1 (F_1' - B_1') + i\omega_2 (F_2' - B_2') \right) \tag{C4}$$

where $F_j'$, $B_j'$ denote the displacement amplitudes of the resonators, which can be expressed as:

$$F_j' = \frac{1 - e^{-iqa}}{2\gamma} \frac{\omega_0^2}{-\omega_j^2 + \omega_0^2} F_j \tag{C5}$$

$$B_j' = \frac{1 - e^{-iqa}}{2\gamma} \frac{\omega_0^2}{-\omega_j^2 + \omega_0^2} B_j \tag{C6}$$

By substituting Eqs. (C5) and (C6) into Eq. (C4) and expressing $\frac{\omega_0^2}{-\omega_j^2+\omega_0^2}$ in terms of $Z_j$, the following expression can be obtained:

$$Z_0^3 A_0\left(1+\frac{k_1}{4\gamma^2 k_p}\right)=\left(Z_1^3\left(F_1-B_1\right)+Z_2^3\left(F_2-B_2\right)\right) \tag{C7}$$

In dimensionless form, Eqs. (C3) and (C7) reduce to Eqs. (16a) and (16b), where $Z_0=\cos\frac{qa}{2}$, $Z_j=\sqrt{\left(1+\frac{\kappa}{4\gamma^2}(1-\frac{\Omega_0^2}{-\Omega_j^2+\Omega_0^2})\right)}\cos\left(\frac{qa}{2}\right)$ $(j=1,2)$.

**Appendix D. Effective stiffness of the damped locally resonant metamaterial defined on the imaginary frequency axis**

This appendix examines the number of frequency bands in which the unit cell shown in Fig. 8 (d) exhibits negative effective stiffness. This number is determined by the solutions of the characteristic equation:

$$-\omega^2+\omega_0^2+2i\zeta\omega_0\omega=0 \tag{D1}$$

When $\zeta_c<\zeta\leq 1$, there exist zero $(\zeta<1)$ or one $(\zeta=1)$ imaginary solution $\omega_i$ for Eq. (D1). Consequently, only one frequency band exhibiting negative effective stiffness exists, as illustrated in Fig. A2 (a) and (b). The upper and lower limits are obtained by solving $k_{\text{eff}}=0$.

For $\zeta>1$, Eq. (D1) yields two imaginary solutions: $\omega_{i1},\omega_{i2}$. These solutions induce singularities (where the effective stiffness approaches infinity) and consequently generate two distinct frequency intervals satisfying $k_{\text{eff}}<0$, as shown in Fig. A2 (c). In this case, the two solutions to $k_{\text{eff}}=0$ correspond to the lower limit of the first band and the upper limit of the second band, while $\omega_{i1},\omega_{i2}$ are the upper limit for the first band and lower limit for the second one, respectively.

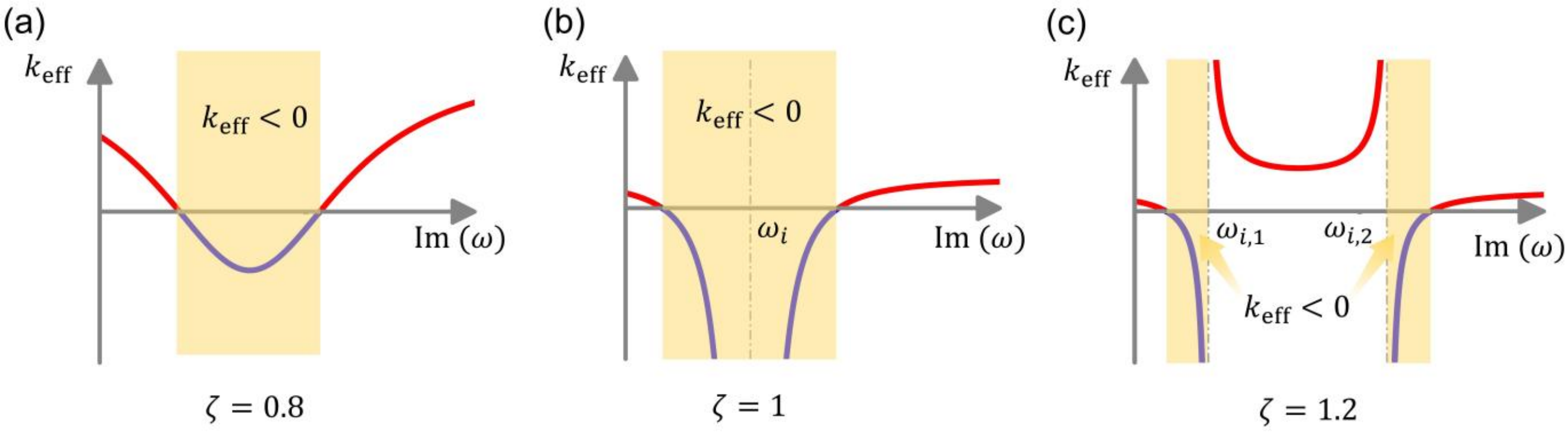

Fig. A2 Variation of the effective stiffness with respect to imaginary-valued frequency for (a) $\zeta = 0.8$, (b) $\zeta = 1$, (c) $\zeta = 1.2$. The parameters are $k_p = k_1 = 1, m_1 = 1, \tan\theta = 1/2$, consistent with those in Fig. 8.